# A Step Towards On-Path Security Function Outsourcing

Jehyun Lee
Trustwave
jehyun.lee@trustwave.com

Min Suk Kang[*]
KAIST
minsukk@kaist.ac.kr

Dinil Mon Divakaran
Trustwave
dinil.divakaran@trustwave.com

Phyo May Thet[*]
A*STAR I²R
phyo_may_thet@i2r.a-star.edu.sg

Videet Singhai[*]
Carnegie Mellon University
vsinghai@andrew.cmu.edu

Jun Seung You[*]
Seoul National University
lucasjsyou0224@snu.ac.kr

## ABSTRACT

Security function outsourcing has witnessed both research and deployment in the recent years. While most existing services take a straight-forward approach of cloud hosting, on-path transit networks (such as ISPs) are increasingly more interested in offering outsourced security services to end users. Recent proposals (e.g., SafeBricks [42] and mbTLS [34]) have made it possible to outsource sensitive security applications to untrusted, arbitrary networks, rendering on-path security function outsourcing more promising than ever. However, to provide on-path security function outsourcing, there is one crucial component that is still missing — a practical end-to-end network protocol. Thus, the discovery and orchestration of multiple capable and willing transit networks for user-requested security functions have only been assumed in many studies without any practical solutions. In this work, we propose Opsec, an end-to-end security-outsourcing protocol that fills this gap and brings us closer to the vision of on-path security function outsourcing. Opsec automatically discovers one or more transit ISPs between a client and a server, and requests user-specified security functions efficiently. When designing Opsec, we prioritize the practicality and applicability of this new end-to-end protocol in the current Internet. Our proof-of-concept implementation of Opsec for web sessions shows that an end user can easily start a new web session with a few clicks of a browser plug-in, to specify a series of security functions of her choice. We show that it is possible to implement such a new end-to-end service model in the current Internet for the majority of the web services without any major changes to the standard protocols (e.g., TCP, TLS, HTTP) and the existing network infrastructure (e.g., ISP's routing primitives).

## 1 INTRODUCTION

Security function outsourcing has been studied [16, 42, 47] in the recent years for its advantages, including cost reduction as well as scalable and resilient provisioning. However, most existing security function outsourcing services are hosted in cloud. Whereas, *on-path* networks, such as ISPs, are increasingly interested in offering outsourced security services to end users. Currently, access ISPs offer security outsourcing services to their enterprises and home subscribers, e.g., [7, 17]. In principle, it is also possible to outsource security functions to non-access, transit networks, such as Tier-1 ISPs, considering recent proposals utilizing emerging hardware primitives. For example, SafeBricks [42] and mbTLS [34] utilize trusted execution environments (TEEs) in commodity hardware to allow outsourcing sensitive applications to untrusted, arbitrary providers.

We envision the next level of outsourcing model — one which enables on-path ISPs to serve security functions (e.g., an IDS, a phishing webpage/email detector [28, 29], a network anomaly detection solution [35, 36], etc.) to *any* demanding traffic passing their networks. Despite the availability of current enabling technologies, outsourcing security services to *arbitrary* on-path networks is not yet ready for prime time. For example, existing deployment scenarios in the state-of-the-art network function outsourcing proposals, e.g., [1, 9, 11, 42], conveniently assume that an outside party has been already determined for outsourcing the functions. The arrangement of one or more arbitrary on-path networks for specific user-requested security functions is not a trivial task. One challenge here is the trust deficit, that users may have with providers, who are seen as curious and potentially malicious entities. Two, in practice, it is difficult to expect the cooperation among on-path networks for forming a security function chain across multiple networks.

Addressing the challenges in outsourcing security functions to arbitrary on-path networks, in this work, we develop and present Opsec (**O**n-**p**ath **sec**urity). Opsec is a novel security outsourcing framework that allows a user to automatically find the security function providers and to establish secure service provisioning channel for her end-to-end sessions. To be more specific, we aim to automate the process of discovering the willing and capable providers who execute the desired security functions and to establish a secure and efficient service provisioning channel only for the selected and verified providers. As a first step, we focus on Web/HTTP(S) sessions in our proof-of-concept design. An example is of a user who wishes to run a secure web gateway and an intrusion detection system (IDS) of his choice for his unsafe web sessions *without* pre-determined secure channel (e.g., VPN connection) with a specific service provider(s). We consolidate our motivation in Section 2, and define the threat model against Opsec in Section 2.2.

Opsec is designed mainly to address three problems. First, when an end user device sends a request for a set of security applications along with a new web session she creates, a new protocol should automatically discover the ISPs that are on the end-to-end path of the session and willing to offer outsourced security services. Also, following the discovery, a micro-contract between an end user and ISPs should be established. Second, during the discovery process, the willing on-path ISPs should be able to efficiently pick up the end-user-generated request messages. Identifying these request messages is similar to finding a needle in a haystack of multi-Gbps transit traffic. Third, a user's privacy should not be tampered with due to the new outsourcing service. To explain, consider encrypted communications, such as in the case with HTTPS. In today's scenario, only three entities have access to the plain-text payloads of an encrypted communication — the user, the server and the contracted security function (e.g., an IDS doing deep packet inspection) via, say, TLS proxy [54]. These conditions should be strictly adhered to even when the security function is outsourced to a third party vendor via on-path ISPs.

The Opsec protocol addresses the above three challenges by exploiting only standard protocols:

- In Section 3, we show that making a security service request to unknown/unspecified on-path ISPs is indeed possible without changes to a user machine and existing protocols. In particular, with Opsec, end users can inform the ISPs on the downstream direction reliably and outsource security applications for the return traffic as well.

- In Section 4, we show that ISPs can efficiently retrieve the Opsec service requests without investing on new routing primitives (e.g., SDNs) and also correctly execute the requested services by only widely available commodity hardware and open-source software stacks.

In Section 5, we evaluate Opsec in the contexts of end-to-end service and ISP networks. To do so, we implement two proof-of-concept prototypes of Opsec, i.e., an end-to-end implementation with a client-side Opsec plug-in and Opsec-box in a commodity machine, and a Docker-based large-scale testbed. We show that an ISP can handle Opsec service efficiently and that a client incurs negligible service arrangement latency. In Section 6, we discuss questions on practicality and security of Opsec.

## 2 MOTIVATION AND THREAT MODEL

We first motivate Opsec with a simple end-to-end example. Then, we describe the rationale of on-path outsourcing, followed by our threat model definition.

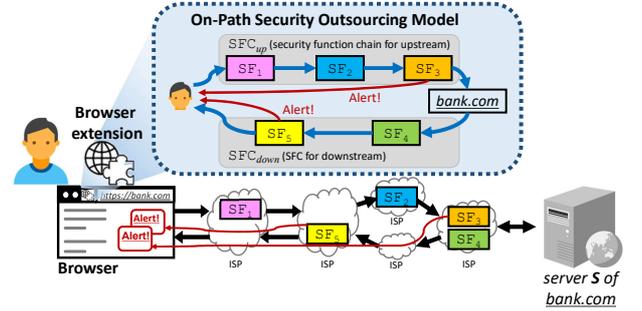

**Figure 1: A motivating example for Opsec.**

### 2.1 An Example

Figure 1 illustrates a motivating example for Opsec. Consider an end user who wishes to establish a new web session with a server $S$ where she needs to be more cautious, like a hyperlink on a social network post. The user wants to apply one or more security functions (e.g., $SF_1, \cdots, SF_5$ in Figure 1) offered by various vendors for her new web session (e.g., a phishing webpage detection solution [29, 31]). The user specifies the desired security functions in the form of two *security function chains* (SFCs): $SFC_{up}$ for the upstream direction (i.e., from the user's machine to the server $S$) and $SFC_{down}$ for the downstream direction (i.e., from $S$ to the user's machine). Each SFC contains a series of standalone security functions of the user's choice. For example, a URL-based phishing detection (e.g., from Symantec) can be chosen for $SFC_{up}$ while malicious script detection (e.g., from FireEye) can be included in $SFC_{down}$. The user can easily configure these SFCs through a dedicated outsourcing agent, e.g., a browser plug-in. After that, when the user establishes a web session with $S$, the browser plug-in makes a preliminary connection for signaling the demand of SFCs to the on-path ISPs. Then, willing on-path ISPs can indicate their capability of executing one or more SFs in the SFCs on the returning packets from $S$ to the client. The final assignment of SFs is automatically done based on the user's policy. After SFC assignment, the browser plug-in proceeds with the main HTTP(S) session to $S$, and all the payloads are sent to the SFs through encrypted secure channels. When an SF wishes to alert the user (e.g., a malicious payload is detected by the IDS), it signals the end user (e.g., with a pop-up security message). The user may also instruct SFs to terminate the web session immediately when certain security policies are violated.

**Outsourcing to On-path ISPs.** Consider connections between a client and a server; a significant gap between the existing outsourcing services using a pre-established network tunnel (e.g., VPN and SD-WAN) and our outsourcing vision, is the support required for unspecified multiple providers in a dynamic way. Restricting to a single and specific provider limits the orchestration of SFCs to only those which are available at the provider. However, outsourcing security functions to multiple service providers that are not on a natural routing path of the end-to-end traffic is a non-trivial problem beyond technical challenges. In order to make the traffic to route through the multiple providers, the service providers must support any form of client-specific and cooperative networking, e.g., the onion routing [12] and VPN chains, to relay the traffic to an arbitrary next-hop provider specified by the client. Notwithstanding technical feasibility, internet-wide cooperation to allow the individual users to specify way-points of an end-to-end path has not been realized.

On the other hand, outsourcing to the on-path providers needs to rely only on the stability of AS-routing and the existence of outsourcing providers on a path. As long as the ISPs provide stable AS-level routing for the lifetime of each connection, the traffic passes the expected ISPs without any additional onion-routing protocol or cooperation of the other ISPs. Thus we argue that the main challenges for on-path outsourcing has challenges are i) in discovering willing and available on-path ISPs, and ii) to provide guarantees of secure and correct service provisioning from the unspecified and untrusted providers.

## 2.2 Threat Model and Applicability Conditions

We now give the threat model against the outsourcing system itself and Opsec's applicability conditions.

**Threat model *against* Opsec.** Opsec aims to make existing security applications/functions more easily accessible to end users. Yet, this new security outsourcing model itself can be a target of new attacks. In particular, we consider new attacks that disrupt or cheat the Opsec security outsourcing model. For *service disruption* attacks, we consider a malicious server $S$ that aims to disable the Opsec security outsourcing so that it can offer malicious web contents (e.g., malware or malicious scripts) to end users more effectively. For *cheating* attacks, we consider on-path malicious ISPs that break a promise to provide the user-requested security function(s). Malicious ISPs may attempt to serve a modified security function to access and store private user data or for economic gain.

**Non-goals.** There are several *non-goals* which we explicitly do not address because they are open problems outside the scope of this work. First, we do not consider volumetric DoS attacks targeting ISP bandwidth [22, 52], as this is not specific to our proposal. Second, we do not consider side-channel attacks against trusted execution environments [6, 30]. Defense mechanisms, e.g., [19, 49] and open-source TEE design [27] can be conducted independently.

**Applicability conditions.** Opsec builds on the state-of-the-art network function outsourcing proposals. Therefore, for Opsec to operate securely against the threat modeled above, the following conditions must hold:

- **[C1]** *Public directory of* SFs. There should exist a public directory of security functions in the market. As long as an end user obtains a tuple of an SF, she can request the SF simply by attaching an SF ID during the service discovery/request phase; see Section 3 for details. Such common, public pool of SFs, including the proprietary solutions of security vendors, is made possible because of the strong confidentiality property of TEEs used in outsourcing; see Poddar et al.'s recent argument [42].

- **[C2]** *Secure payment channel.* We assume a *secure payment channel* with the security functions SFs that runs in TEEs. When an end user establishes a channel with an SF, neither one of them can cheat (e.g., refuse to pay) because the payment process is securely escrowed by a TEE-protected payment logic, often combined with smart contracts. There has been a growing interest in the construction of such TEE-based micropayment channels in the last few years (viz., Kosto [10], SPOC [24]) and it is orthogonal to our Opsec outsourcing framework.

- **[C3]** *Verifiable resource accounting.* Another underlying required primitive to guarantee verifiable resource accounting for accountable billing system. This ensures that the final billing issued by a SF indeed reflects the correct measurement of the resource used for the function execution, not an inflated billing. This highly desirable primitive has been actively studied in the last few years [1, 18, 51].

- **[C4]** *TEE infrastructure.* To enable all the TEE-based security primitives and constructions, there must exist a highly available public-key infrastructure for the attestation of TEE hardware; e.g., Intel Attestation Service [21].

In addition to these mandatory conditions, for a clear presentation of our design, we assume there exists at least one ISP on an end-to-end path willing to provide any SF requested by end users.

## 3 SECURITY SERVICE OUTSOURCING TO ON-PATH ISPs

Opsec users wish to outsource security services of their choices to one or more *unspecified* on-path ISPs. In this section, we focus on the technical challenges for discovering willing and capable ISPs on the path between a user and a server. Here, we present our end-to-end service request protocol design between a client and on-path ISPs, which we denote as the *Opsec* protocol.

We first clarify the five properties for the on-path security service outsourcing in the clients.

### 3.1 Desired Properties for Outsourcing Security Functions to Unspecified ISPs

- **[P1]** *Expressive security service requests.* An end user should be able to request an arbitrary list of security functions in the form of a function chain. This is challenging because the requested security functions should be provided even by external providers *without* any existing arrangement of security services.

- **[P2]** *Session selectivity.* A user should be able to select which end-to-end session should receive the requested security services. For example, a user may want to run some DPI functions only for specific web sessions.

- **[P3]** *Legacy client support.* Opsec should not modify the standards of existing network protocols (e.g., IP, TCP, and TLS) and their implementations at both client/server sides.

- **[P4]** *Service offering by one-direction ISPs.* Compared to requesting services to ISPs on both upstream and downstream paths, doing so to ISPs only on one-direction path is much harder mainly because communication between clients, upstream one-direction ISPs, and downstream one-direction ISPs is not trivial. For example, it is hard for a client to send the service request messages to the one-direction ISPs on the downstream path. Also, sending handshake replies and alert messages from an upstream one-direction ISP to a client is challenging because the ISP is not on the downstream path.

- **[P5]** *No server-side support.* We argue that servers should not actively cooperate with their clients for the Opsec service (e.g., delivering Opsec messages to downstream networks on behalf of the clients) because it requires significant changes to the current server protocols and implementations. Achieving such coordination in practice would be extremely challenging (if not impossible) as it requires global-scale coordination (and thus software upgrades) of servers and their protocols.

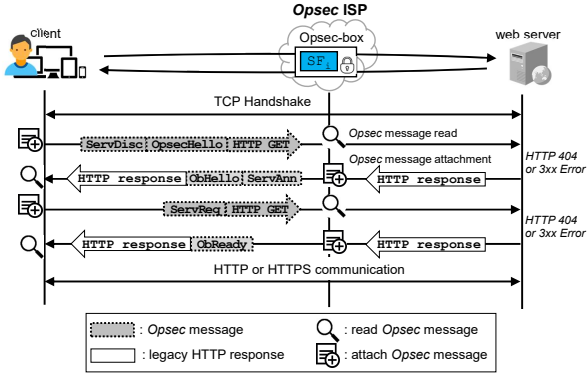

Figure 2: Opsec service request protocol over HTTP.

| 0 | 1 | 2 | 3 | 4 | 5 | 6 | 7 (octect) |
|---|---|---|---|---|---|---|---|
| Opsec Start Tag | | | | Message type | | Opsec-box ID | |
| Opsec-box ID | | Message Length | | | | Message | |
| Message ... | | | | Opsec End Tag | | | |

Figure 3: Opsec message format on a HTTP GET request and response.

### 3.2 Security Service Request Protocol

At a high level, the Opsec protocol is composed of two end-to-end rounds: the first for discovering the existence of Opsec services provisioned by willing and available ISPs on the path, which we call *Opsec ISP*, and the second for establishing a secure channel with a service instance that understands the protocol, which we call *Opsec-box*, operated by the discovered Opsec ISP. To that end, we propose a HTTP-based protocol, which convey the messages for secure channel establishment and outsourcing negotiation upon HTTP GET requests and responses. Figure 2 presents the design of Opsec protocol. In the figure, we depict an Opsec-box assuming that it is on *both* the upstream and downstream directions between a client and a server. Later in Section 3.3, we present how Opsec relaxes this assumption. This Opsec handshake step requires only up to 3.5 round-trip time (RTT) additional latency including 1.5 round of TCP connection. Once this one-time handshake is completed, packets are processed by these ISPs for the arranged security functions.

Note that this key sharing and secure channel design is inspired by the standard TLS v1.2 and mbTLS [34]; yet, unlike mbTLS, it does not require any changes to the TLS protocol and TLS implementations being used by the clients. The session key creation and secure key sharing with asymmetric keys and symmetric cipher in Opsec system follow the standard algorithms and implementations of a fixed set of cipher specifications (e.g., ECDH_RSA_AES256-SHA), thereby avoiding the negotiation of cipher suites during handshake.

The Opsec protocol satisfies all three properties **[P1]** to **[P3]**. First, for **[P1]**, we define a generic Opsec message format shown in Figure 3 so that users can specify the required security services (with their unique ID). Also an Opsec-box can recognize Opsec messages on HTTP GET messages, which may not always be at a static position, with Opsec start/end tags. Second, for **[P2]**, we choose HTTP(S) sessions as the granularity of the Opsec services; that is, a user specifies distinct security services for each HTTP(S) session. Third, for **[P3]**, we make sure that no modifications of clients or servers (and their existing protocols) should be for Opsec.

We summarize six types of Opsec messages found in Figure 2 in order of appearance:

1. OpsecHello (Opsec hello): This message works as a beacon to discover any Opsec-boxes that understand the Opsec protocol. It includes the client's public key and the client's random nonce for secure symmetric key generation and the encryption of the return messages.

2. ServDisc (service discovery): This message carries a list of unique IDs for the requested security functions. The integrity of this list is guaranteed as the client can check the signed hash of this message later in the RbReady message.

3. ObHello (Opsec-box hello): This message carries the freshly-generated public key of the Opsec-box for the subsequent secure session establishment. Unlike the TLS public keys of the end servers, the Opsec-box public keys are authenticated with the external certificate authority of the trusted execution environment platforms (e.g., IAS of Intel SGX [21]). The message thus also contains the attestation quote for the Opsec-box boilerplate code that later loads and executes the requested functions.

4. ServAnn (service announcement): Each Opsec-box replies with a list of the hashes of available security functions.

5. ServReq (service request): The final message from a client is to confirm the allocation of specific security functions at specific Opsec-boxes. This message is followed by the freshly-generated master-secrets to the selected Opsec-boxes, which will later be used to generate the per-hop session keys; see Section 4. The master-secrets are encrypted with the public key of that particular Opsec-box.

6. ObReady: This is the final message from each Opsec-box that confirms that the Opsec-box is ready to perform the Opsec service. It also carries the signed hash of ServDisc and ServReq for integrity check of the list of requested security functions against malicious modification by ISPs in the middle and repudiation of the requested functions.

The handshake requires a response from the web server, irrespective of the response type, to allow the Opsec-boxes to attach their responses for OpsecHello and ObReady, i.e., ServAnn and ObReady. We confirm the expected behavior of a web server shown in Figure 2 in our experiments for the Alexa top-300K web sites ranked in 2019 [2]. That is, all the accessible web servers in the list return a HTTP response when our test client sends a few hundred bytes long GET requests with the Opsec messages. In detail, more than 90% of the web servers return a standard HTTP response for sending a HTTP error code, e.g., 3xx (63.40%) or 4xx (29.40%). 6.34% of the responses return with a non-standard error code. Yet, a malicious server that is aware of Opsec protocol may attempt to exploit the protocol. We discuss the possible attack scenarios by the Opsec-aware servers in Section 6.

## 3.3 Reflection for One-Direction ISPs

We now consider the one-direction ISPs and show how Opsec achieves all the previous properties **[P1]**–**[P5]** for the one-direction ISP cases. The idea is to *reflect* the Opsec-specific additional information sent by clients at the servers so that downstream ISPs obtain them.

The reflection should be carefully designed though because we need to make sure that the following two types of Opsec-specific information in two different layers arrive at the downstream ISPs: the Opsec messages in HTTP (see Figure 2) and the TCP port numbers required for indicating Opsec service traffic. We will describe how Opsec manages TCP port numbers to enable Opsec ISPs to identify Opsec traffic efficiently in Section 4.2. Moreover, all the reflection should be performed without direct support from the servers.

*3.3.1 HTTP URL reflection.* The first reflection technique that we use is a feature that can be found in the majority of HTTP web server implementations; see another system that exploits this reflection feature [33]. Web servers usually return back the requested URL from a client in the reply messages when they attempt to redirect a client to a specific domain, port, or a valid entrance URL. Once the server replies to the HTTP request with the reflected URL, the downstream Opsec-boxes receive the Opsec messages. Note that this URL reflection can also be used to deliver messages (e.g., warning messages for phishing attempts) from upstream Opsec-boxes all the way back to the clients.

Opsec clients request a non-existent HTTP URL that includes the desired Opsec message to end web servers, and we commonly see the two different URL-reflection behaviors:

1. *URL reflected in redirection messages.* As explained above, on receiving requests to the HTTP port, many web servers redirect the clients to the corresponding HTTPS servers, and during this process they reflect (i.e., return) the requested URL (i.e., with the *path* following the domain name). This is because browsers often make HTTP connections by default[*], and thus they need to be redirected to the HTTPS port by the servers. Opsec leverages this redirection to achieve the URL reflection.

2. *URL reflected in error messages.* In some minor cases, web servers respond with error messages but still reflect the requested URLs in their error messages.

Figure 4 shows how Opsec exploits the HTTP URL reflection technique to facilitate communications between upstream entities and downstream entities. The Opsec messages from a client and the Opsec responses appended by upstream Opsec-boxes are *reflected* by a web server. Then, downstream Opsec-boxes and the client receives the Opsec messages and the Opsec responses, respectively.

## 4 SECURE AND EFFICIENT EXECUTION OF SECURITY SERVICES AT ISPS

We have seen in Section 3 that the Opsec protocol addresses many requirements for security service outsourcing to on-path ISPs, from the end users' perspective. However, Opsec is still incomplete because there still remain several non-trivial technical challenges to execute the requested security functions in transit ISP networks. In this section, we answer (1) how an ISP efficiently identifies the Opsec sessions and redirects them with only legacy switching equipment, and (2) how the Opsec-boxes provide confidentiality and integrity for the required function executions.

[*]In the case of Firefox, redirection is observed when a domain is visited for the first time within a browser session.

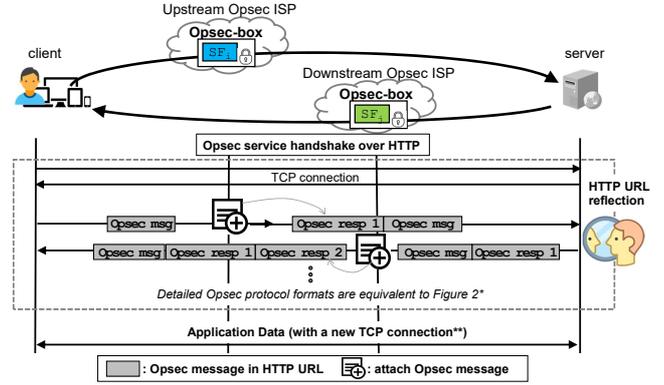

Figure 4: Reflection of Opsec messages using HTTP URL reflection (in Opsec) for supporting ISPs that are only on one-direction path.

We first summarize two desired properties for efficient, trusted execution of security services and then propose the traffic handling mechanism in Opsec to achieve these two properties.

## 4.1 Desired Properties for Secure and Efficient Execution in ISPs

- **[P6]** *Flow redirection in legacy networks.* When Opsec packets travel through an ISP that is willing and capable of providing the requested security functions, the packets must be redirected to the Opsec-boxes with no discernible added delays. However, it is quite challenging to achieve it with *legacy border routers* in ISPs, where dynamic rerouting capability is limited (e.g., no programmable switching capability available). Our Opsec protocol must make such redirection of Opsec packets perfectly backward compatible to legacy border routers.

- **[P7]** *Trusted execution of security services.* When Opsec packets are processed by an Opsec-box, Opsec clients should be able to validate whether the correct security functions are executed.

## 4.2 Efficient Opsec traffic Redirection in ISPs

For the property **[P6]**, legacy edge routers of an Opsec ISP should i) distinguish Opsec packets from non-Opsec ones in real-time and ii) redirect Opsec packets to the Opsec-boxes in its ISP network. Note that this can be accomplished easily when all edge routers are capable to read and understand a special protocol, like the network service header of SDN. To avoid such a forklift upgrade of the ISP backbone and enable easier adoption of Opsec, however, we use only traditional forwarding and tunneling mechanisms in the backbone.

Let us first explain our basic idea of how Opsec ISPs identify and redirect Opsec packets with only legacy edge routers, as shown in Figure 5. For this, we exploit only the *TCP port numbers* (both source and destination). In particular, Opsec users choose a destination port number $p^*$ from a static, predefined set of *Opsec port numbers*, $\mathcal{P}$, as an indication for the Opsec service, instead of using the listening port number $p_s$ of the server (e.g., 443 for HTTPS servers, 80 for HTTP servers).[†] For $\mathcal{P}$, we can choose a small number of TCP port numbers that are not yet used for popular services so as to minimize false positives as well as not on the range of usual source ports and thereby minimize unnecessary redirection of non-Opsec packets to Opsec-boxes. Furthermore, for indicating the flow to a destination port $p_s$ with $p^*$, a set of ports $\mathcal{P}$ is bond to a $p_s$, (e.g., 7443 and 8443 to 443). Then, as shown in Figure 5, any traditional edge routers can easily distinguish the Opsec and non-Opsec packets and redirect all Opsec packets through the pre-defined MPLS tunnels to the Opsec-boxes.

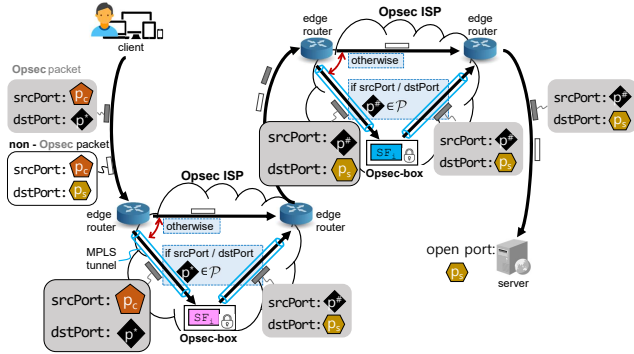

Figure 5: Example of TCP port re-arrangement in Opsec. This simple TCP port re-arrangement enables efficient Opsec packet redirection at legacy edge routers.

Although this TCP-port based solution is simple and effective in identifying and redirecting Opsec packets at legacy edge routers, care must be taken to ensure that this does *not break* end-to-end TCP sessions. We outline four problems of the TCP-port based solution and describe how we address them:

1. Since a destination port number $p^*$, which is different from the listening port of the server $p_s$, is used, the server would drop Opsec packets if the Opsec-box simply forward the packets after their executions. We address this problem by rearranging the chosen Opsec destination port $p^*$ back to $p_s$ before the packets leave the Opsec-box. an Opsec-box obtains $p_s$ from a pre-defined mapping between a $\mathcal{P}$ and a $p_s$. This way, Opsec packets arrive at the server with a right destination port number.

2. As a result of the destination port number re-arrangement, however, Opsec packets now lose the indication of the Opsec service (i.e., $p^*$), and thus no other Opsec ISPs after the first one can identify them. This is particularly important when the requested security function chain is long and thus more than one ISPs should execute security functions. To make the Opsec packets easily distinguishable even after visiting the first Opsec-box, we replace the source port number of the Opsec packets leaving the first ISP with $p^\# \in \mathcal{P}$. From the second Opsec ISP onward, their edge routers can easily distinguish Opsec packets since they have $p^\# \in \mathcal{P}$ as their TCP source port; see Figure 5. We replace $p^*$ to $p^\#$ because of the next problem.

[†]Note that we cannot use the source port set by the client because the source port is often changed by a NAT.

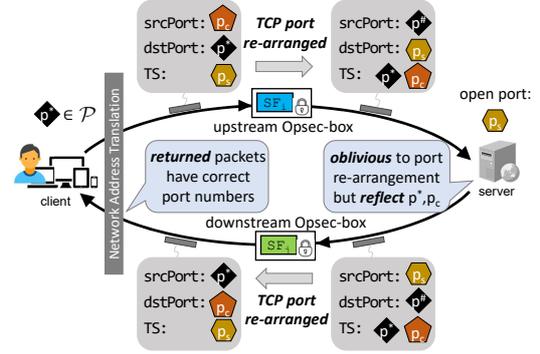

Figure 6: TCP port re-arrangement together with the TCP timestamp (TS) reflection. This careful port and the timestamp assignment enables efficient packet redirection in ISPs on both directions.

3. When multiple clients behind a NAT try to have Opsec serviced connections to the same server with a same destination port $p^*$, replacing the source port numbers with the destination port $p^*$ at the first Opsec ISP leads source port conflict at following Opsec ISPs and the server. To avoid this source port conflict, the first Opsec ISP assigns new source ports $p^\# \in \mathcal{P}$ unique within the flows sharing the same source and destination IP addresses.

4. When Opsec packets return back to the clients, they must have the original source port number $p_c$ and the destination port number $p^*$ because otherwise the clients (or NATs in front of them) would drop the packets. To address this, all the Opsec-boxes that re-arrange the TCP port numbers of Opsec packets in the upstream direction should restore the port numbers for the Opsec packets towards the client.

Note that the above port-number re-arrangement may lead to *false positives* when non-Opsec packets happen to use the port numbers in $\mathcal{P}$. Yet, such false positives create neither any service disconnection nor degradation to the affected non-Opsec packets because when the client sends the application data, the Opsec-boxes will observe no Opsec record, and henceforth will only forward the packets without any processing or breaking the session semantics. Therefore the session will continue as normal.

*4.2.1 TCP timestamp reflection.* As we addressed above, the packets coming back to the clients (or a NAT in front of them) should use $p_c$ (i.e., the original source port number) as their destination port number and $p^*$ (i.e., the original destination port number) as their source port number. This TCP port re-arrangement can be easily done by ISPs on both the upstream and downstream paths but not by one-direction ISPs because one-direction ISPs on the downstream path are not aware of $p^*$ and $p_c$. We exploit the timestamp option of TCP headers (which is reflected by TCP servers) to convey $p^*$ and $p_c$ values all the way to the ISPs in the downstream direction. Figure 6 illustrates how $p^*$ and $p_c$ are conveyed to the downstream Opsec-boxes with keeping correct port numbers for

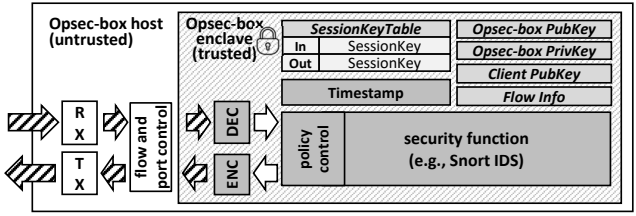

**Figure 7: Opsec-box design with a trusted execution environment.**

both of the server and the client. In this way, the final destination port number of the return packets is set to $p_c$.[‡]

Note that although we manipulate the TCP timestamp for a non-standard purpose, it is unlikely to cause disruptions. The two standard use cases of TCP timestamp [5] include the troubleshooting of the Internet connectivity and the protection enhancement in high-speed networks, which will not be used by most individuals.[§]

### 4.3 Confidential and Correct Execution of Security Functions

**Trusted execution environment for security functions.** To enable the trusted execution of the outsourced security functions (i.e., **[P7]**) at arbitrary ISPs, we exploit the trusted execution environment (TEE) that is widely available in the commodity CPUs these days (e.g., Intel SGX [32] and ARM TrustZone [3]). Figure 7 illustrates the high-level architecture of the TEE-based security function executions. Note that our design is inspired by several recent projects, such as ClickSGX [9] and SafeBricks [42], to protect the network functions with the TEE's confidentiality and integrity guarantees. In addition to the network functions, we place the asymmetric key pairs, session keys, flow information, and packet payloads within the protected memory region called an enclave.

**TLS key sharing and inter-ISP secure channel.** When an Opsec flow uses TLS, e.g., HTTPS, the security services which need to inspect the plain text of the application layer payload, e.g., malware detection, require a TLS session key to decrypt the TLS payload.

Opsec achieves the TLS key sharing and inter-ISP secure channel by establishing per-hop secure channel and handing over the master secret of the TLS session between a client and its server to the Opsec-box which is an immediate hop from the server on ServReq. As we noted at Section 3.1, ServReq is protected by asymmetric cipher and only readable by a specific Opsec-box. If the flow served by Opsec is a plain text channel, e.g., HTTP, the application payload on the last hop is sent in plain text for compatibility to the server. The hops between Opsec-boxes are protected by the per-hop secure channel irrespective of TLS between the client and server. Unlikely the standard TLS, Opsec-box does not create its master secrets by itself. This is designed for expressive service chain orchestration that only the Opsec-box selected by a client participate to the service chains. More specifically, Opsec client creates master secrets following the fixed cipher specification with Opsec-box's public key and random number gathered from ObHello for each hop between a client to a server, and sends the master secrets over ServReq messages to each corresponding Opsec-box.

---

[‡]TCP Echo option would have been an ideal feature for this purpose; however, it is obsoleted by the timestamp option [14].
[§]If network operators of upstream networks wish to utilize the TCP timestamp standard features, they can easily overwrite TCP timestamp set by Opsec clients.

## 5 FEASIBILITY EVALUATION

In this section, we evaluate feasibility of Opsec with two proof-of-concept implementation of Opsec, an end-to-end implementation, and a Docker-based large-scale testbed. Through the end-to-end experiments, we evaluate additional latency by one-time service arrangement and processing latency by Opsec-box. We also show scalability of Opsec service in an ISP network and feasibility of congestion control induced by Opsec traffic in the large-scale testbed.

### 5.1 End-to-end Opsec Implementation

We evaluate the performance of an Opsec-box in a simple end-to-end network setup with a browser client and a web server. Particularly, we measure the latency and throughput performance of the end-to-end Opsec sessions (i.e., an Opsec-box sits on the client-server path) in comparison with non-Opsec sessions.

**Opsec browser plug-in Design and Implementation** We implement a proof-of-concept of the Opsec browser plug-in as a Google Chrome browser extension. The plug-in, once activated, grabs the web requests generated by a user before making a TCP connection, and performs an Opsec handshake for each web request towards the target web server. The browser plug-in then talks to a background daemon that conducts cryptographic operations and TCP port re-arrangement on the session traffic. The daemon executes the cryptography operations (e.g., OpenSSL [38]) for secure Opsec handshake and key generation. After an Opsec handshake, the plug-in redirects the original web request with the Opsec service TCP destination port number $p^* \in \mathcal{P}$. This proof-of-concept plug-in is implemented in JavaScript with 63 source lines of code (SLoC), and the Opsec daemon is a C++ application with 850 SLoC with OpenSSL-1.1.1 and netfilter-queue 1.0.2.2 library. To support non-browser HTTP(S) connections, one can also consider the full-implementation of the client-side functionalities in the background daemon (i.e., without the need of a browser plug-in).

**Opsec-box Design and Implementation** We design and implement an Opsec-box to execute Snort [45], which is a widely used intrusion detection system in virtualized environments. We use Intel SGX and Intel DPDK [44] for the trusted execution environment (TEE) and line-rate packet processing, respectively. We base our Opsec implementation on an open-source Snort-SGX-DPDK project [25] that relies on the Graphene project [53].

Our Opsec-box implementation consists of two parts: the Opsec handshake (Section 3.2) and the TCP port re-arrangement and service flow control (Section 4.2). To implement the Opsec handshake and packet enc/decryption in the reference Snort-SGX-DPDK system, we modified 840 SLoC in snort.cc in Snort. The TCP port re-arrangement and flow control are implemented with only 188 SLoC in daq_dpdk.c. We use: Ubuntu 16.04.6 LTS, Intel SGX Driver v2.5 release 0b76a7c, Intel DPDK driver 17.08, Snort3 build 239, Graphene-SGX [53], LibDAQ v2.2.1, and OpenSSL v1.1.0 [38] (as part of the Snort-SGX-DPDK project [25]).

**Experiment Setup.** We built a simple testbed by placing an Opsec-box between a client and a legacy HTTPS web server. The Opsec-box is equipped with Intel® i5-8400 CPU (6-core of 4.3 GHz), 8 GB of main memory, and Intel® X550-T2 10 Gbps NIC. The client and the Opsec-box are directly connected to each other with an STP CAT6 cable.

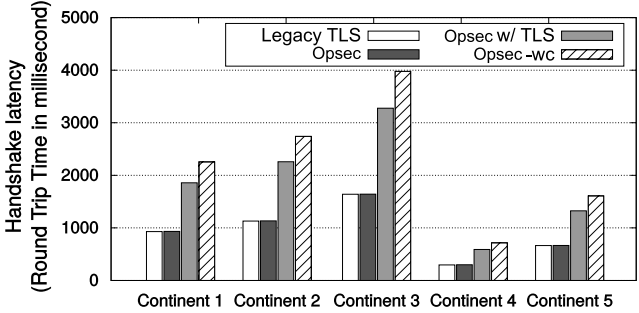

Figure 8: Handshake latency comparison of legacy TLS, Opsec, and Opsec with TLS to five web servers in five different continents. We hide the continent names for double-blind submission.

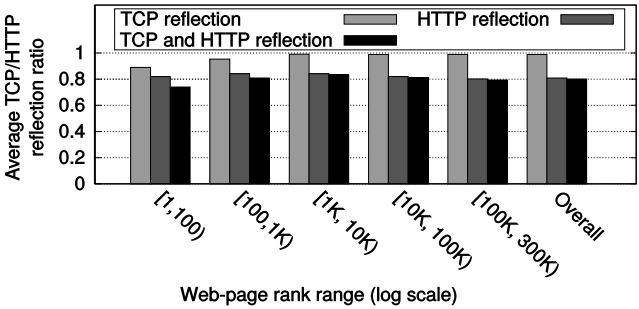

Figure 9: Ratio of servers that support TCP timestamp reflection, HTTP URL reflection, and both.

### 5.2 End-to-end Handshake Latency

We evaluate the end-to-end latency of the Opsec handshake. We measure the handshake latency between a client and five web servers in five different continents and compare them with the TCP/TLS handshake of the legacy TLS. The end-to-end Opsec handshake latency is shown to be dominated by the round-trip time (RTT) between the client and the server. That is, the computational overhead of the Opsec handshake is negligible compared to the packet travel times. Figure 8 shows the difference between legacy TLS and Opsec. In practice, there is no discernible difference in the latency between legacy TLS and Opsec in all five cases, showing that the computation overhead of the Opsec-boxes is small. The HTTPS connections with Opsec establish TLS session after Opsec handshake, that require additional 3.5 RTT for TLS. Opsec-wc denotes the worst case of the handshake, where web servers disconnect TCP sessions after one HTTP response. Thus, Opsec-wc requires 1.5 more RTTs for a TCP handshake.

### 5.3 HTTP URL and TCP timestamp Reflection

We test the two reflection behaviors in the Alexa top-300K web sites ranked in 2019 [2]. The result is promising. Figure 9 shows how widely the HTTP URL reflection introduced in Section 3.2 and TCP timestamp reflection behaviors are available in the current Internet. In all the website, TCP timestamp reflection seems to be universally supported. This is because the TCP timestamp reflection is mandatory in the standard [5]. HTTP URL reflection is also supported by the majority (e.g., >80% in all the ranking ranges) of existing web servers, although it is not a mandatory behavior. Specifically, 71% of URL reflections are the result of the HTTPS redirection and 10% are from the error messages when the HTTPS redirection is not supported. Overall, $\approx 80\%$ of existing web servers support both the TCP timestamp and HTTP URL reflections, making our Opsec service available for the majority of web services today.

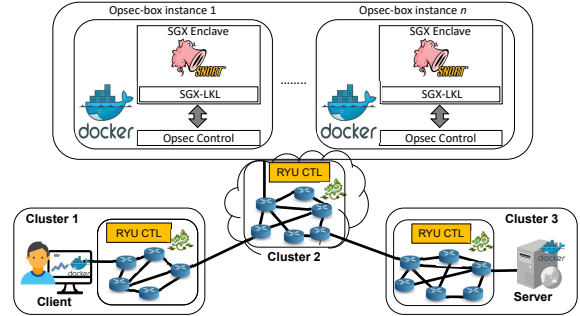

Figure 10: Implementation overview of our container-based large-scale experiment of end-to-end Opsec service.

### 5.4 Scaling up Opsec Service in ISP Networks

We show that ISPs can operate the Opsec service in a scalable manner with virtualized Opsec boxes and traffic engineering using only existing routing primitives. Opsec, like other virtualized function platforms [39, 50], is able to scale up its total processing capacity by increasing the service instances in its computing cluster infrastructure. We show that we can adjust the number of Opsec-box instances dynamically in a network function virtualization testbed.

**Container-based testbed.** We construct a Docker container-based testbed on three server clusters with Intel Xeon CPU E5-2620 (32-core of 2.10GHz), 64GB of main memory, and one Intel X540-AT2 10Gbps network interface card. As shown in Figure 10, we place Ubuntu-16.04 Docker instances for a client (Opsec agent and HTTPerf), a server (Apache Server) and the Opsec-boxes (Snort over Simulated SGX) on three different clusters. Each cluster has Open vSwitch (OVS) v2.5.4 [37] for network communication between Docker instances. OVSs on each cluster are controlled by RYU SDN controller [46] via OpenFlow 1.3 protocol. To implement the Opsec-box instance, we install the Opsec control binary file and SGX-LKL [43], a library OS that enables to run Linux applications inside SGX enclaves. We can implement the Snort-2.9.12 inside an enclave with only 104 SLoC modification of snort.c including Opsec flow control. We assign 3 cores and 5 GB of memory to each Opsec-box Docker instance.

**Latency performance independent of traffic load.** As the traffic load to the Opsec functions increases, the functions become overloaded, and users start to experience increased end-to-end latency. We show that the end-to-end latency can be well controlled even when the offered traffic load significantly varies. The idea is simple. Opsec service assigns new Opsec Snort service instances for new Opsec flows that would exceed a pre-defined capacity of the Snort instances. RYU controller distributes Opsec flows and balances the load.

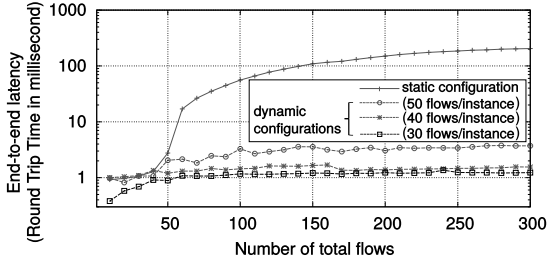

Figure 11: End-to-end latency of Opsec service in a large-scale experiment for varying numbers of Opsec flows.

Table 1: Five ISP topologies and test traffic matrices $\mathcal{T}$ used in the evaluation of ISP internal routing of Opsec traffic.

| Network topology | # of nodes | # of links | Link cap. | # of flows in $\mathcal{T}$ |
|---|---|---|---|---|
| GTS CE | 148 | 386 | 2,000 | 1,500 |
| Telcove (Level 3) | 70 | 140 | 13,200 | 1,040 |
| Columbus Net. | 69 | 170 | 10,000 | 1,008 |
| Missouri Net. | 66 | 166 | 10,000 | 915 |
| Forthnet | 61 | 124 | 6,000 | 770 |

Figure 11 shows how the end-to-end latency measured by clients changes for varying offered load to the Opsec-boxes. We test one static configuration (see a solid line) and three different dynamic provisioning configurations (see three dashed lines) with different threshold values. As the number of total flows increases, the static configuration that uses a single Opsec instance quickly causes significant delays (note that the Y-axis is in log-scale). On the contrary, we dynamically add new Opsec instances when the average number of flows for per-instance exceeds a certain threshold (e.g., 30, 40, and 50 flows per instance) to handle the increasing flows without much performance degradation. The three dashed lines indeed show that the end-to-end latency stops increasing even when the number of flows keeps growing. Particularly, when the threshold is set properly (e.g., 30 flows per instance in this experiment), we barely see any degradation in end-to-end latency regardless of the total number of incoming flows.

### 5.5 Avoiding Congestion Induced by Opsec

Opsec traffic may be treated differently within an ISP and it may cause unexpected congestion. Particularly, when a large number of Opsec sessions are redirected to travel through the Opsec-boxes, congestion is expected near the computing cluster that hosts the Opsec services. We show that, however, such congestion induced by Opsec traffic can be easily avoided by a simple traffic engineering mechanism with legacy traffic tunneling mechanisms, such as Multi-Protocol Label Switching (MPLS). We test our simple MPLS-based tunneling traffic engineering in various, realistic ISP internal network topologies and traffic conditions.

We evaluate traffic engineering in five ISP network topologies (see the details in Table 1) gathered from Topology Zoo [23]. We select the five largest (in terms of the number of nodes) commercial

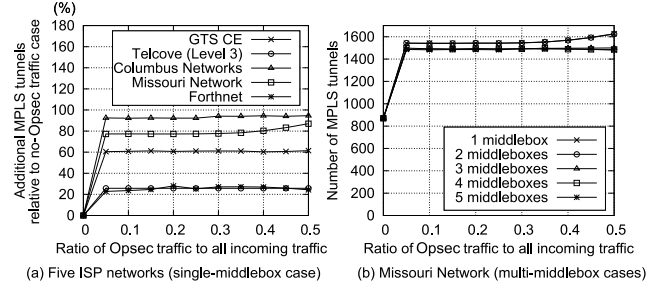

(a) Five ISP networks (single-middlebox case)   (b) Missouri Network (multi-middlebox cases)

Figure 12: Measured routing cost changes (in terms of the number of required MPLS tunnels) for varying the ratio of Opsec traffic to all incoming traffic. (a) Additional MPLS tunnels relative to no-Opsec traffic case for five ISPs. (b) Number of MPLS tunnels for the case of Missouri Network.

networks from the Topology Zoo data set. For a realistic evaluation, we use gravity-random distributed traffic matrix $\mathcal{T}$ [56].

We compute the optimal-cost routing while increasing the ratio of Opsec traffic from 0% to 50% of the traffic by expressing the problem as well-known in integer linear programming (ILP) formulas[¶] and construct distinct end-to-end paths to implement the routing solution with MPLS. We assume that the cost of each network hop and set the link capacity between the nodes are uniform to all the links to highlight the effect of MPLS based routing.

Figure 12 shows the routing cost changes in terms of the number of MPLS tunnels in our evaluation for varying traffic load to the above five ISPs. Particularly, we increase the ratio of Opsec traffic to all incoming traffic to these ISPs. Here, we assume that all the ingress-egress pairs of these ISPs already use one MPLS tunnel for traffic at each direction. We focus on the increase of MPLS tunnels to accommodate the increasing Opsec traffic that should be redirected towards one of Opsec-box inside the ISPs. Figure 12(a) shows that as we turn more incoming traffic into Opsec traffic, all five ISPs should create additional 23%-95% of MPLS tunnels to support the redirection while avoiding traffic congestion. As we increase the ratio of Opsec traffic, we expect more MPLS tunnels; yet, the increase is only about 3-10 percentage points after the first increase of the tunnels. This shows that the routing cost is, in fact, quite manageable in typical ISPs. Figure 12(b) shows how this routing cost can be further reduced when we deploy Opsec-boxes throughout the ISP networks in a distributed manner. In the case of Missouri Network Alliance, as we use more Opsec-boxes, the routing cost (i.e., the number of MPLS tunnels) decreases because the distributed Opsec-boxes lower the risk of congestion in the ISP network and thus removes restrictions for traffic redirection.

## 6 DISCUSSIONS

We discuss Opsec's practicality and threats it could face.

### 6.1 Common Questions

- *Why would on-path ISPs be interested in Opsec?* Access ISPs have *already* started to offer security outsourcing as a value-added

---
[¶]We solve the problem by an ILP solver, i.e., IBM CPLEX (https://www.ibm.com/analytics/cplex-optimizer). Refer to the detailed formulas in Appendix A to implement the routing solution with MPLS.

- service to their customers; see AT&T's Global Security Gateway Service [4] or Fusion Connect's SD-WAN security solutions [17]. Transit ISPs[||], however, have not been able to take advantage of this opportunity since they do not have direct subscribers to sell their outsourcing services. Opsec enables transit ISPs to *monetize their topological advantage* (i.e., naturally being on the paths of many Internet traffic flows) by providing security solutions as a service to many end users.

- *What are the user-side benefits of outsourcing to on-path ISPs, compared to outsourcing to clouds via VPN?* In addition to the main advantages of Opsec, outsourcing to on-path ISPs simplifies the end-to-end communications. That is, the same inter-domain paths are used when on-path ISPs perform outsourced tasks. Thus, end users experience no increase of end-to-end routes.

- *Would ISPs have enough compute resources for Opsec?* Many ISPs already have such general-purpose compute resources for value-added services [15, 55], and they can be easily repurposed to host Opsec because it only needs commodity, general-purpose compute clusters with TEE support. Scaling up of the deployment and operation of such general-purpose compute clusters is a well-studied topic in the last decade [15, 42, 47, 55].

- *Can we deploy Opsec incrementally?* At an early deployment stage, it requires *only* a small number of large ISPs, such as Tier-1 ASes, to support Opsec to cover a large portion of the Internet. Gradually, Opsec can be adopted by more ISPs so that the outsourced security services can be performed in a more distributed and sustainable manner. That said, we foresee non-trivial difficulties of forming a consensus in the network operator's community (e.g., NANOG) and standards organizations (e.g., IETF) because this change would involve multiple stakeholders across different geopolitical regimes. In this work, we focus on a technical means to this novel security outsourcing model, and hope to spark new discussions on this vision.

### 6.2 Security Analysis

Upon Opsec operation, adversaries may attempt to disrupt Opsec components, i.e., client, server, and ISPs on the path, to exploit Opsec protocol or disrupt the system itself. However, Opsec design is robust against such threats.

**Dishonest/Malicious ISPs.** A dishonest on-path ISP may attempt to deceive a client by overcharging or providing incorrect service. Opsec system guarantees the integrity of the execution through trusted execution environments (TEEs) and a remote attestation infrastructure, such as Intel Attestation Service [21]. On the other hand, an ISP may disrupt Opsec protocol or Opsec service traffic without regarding loss of connection. However, these disruptions are highly expensive for an ISP compared to the potential benefits from the services offered.

**Fake-Opsec ISP.** A malicious server with TEE may act as a fake Opsec ISP for an end-point; however, the server also has to follow the Opsec service request process since it has to have a genuine TEE. Disruption of a service by a fake-Opsec ISP is hardly beneficial;

rather it prevents a potential victim from accessing its malicious server.

We discuss these potential threats further in Appendix B.

## 7 RELATED WORK

We share the same motivation for outsourced security services for home networks presented in a position paper by Feamster in 2010 [16]; i.e., instead of individual home networks, external parties (with more resources) should execute security functions. Opsec sits at the confluence of three different lines of work: study and design of general outsourcing framework for network functions, flexible function virtualization for the mitigation of sophisticated network threats, and verifiable network function executions in third-party systems.

**Network function outsourcing.** Researchers have long envisioned outsourcing network functions from individual networks to more centralized networks (e.g., cloud infrastructure) for cost reduction and easier management. APLOMB [47] is one of the earlier proposals that motivates the advantages of network function outsourcing for enterprise networks with the implementation of a cloud-based outsourcing system. More recently, Embark [26] augments APLOMB architecture to widen the range of outsourced functionalities to DPI from traditional flow-based network functions.

**Network function virtualization for security applications.** In a different line of work, researchers have developed a much flexible, scalable network function virtualization (NFV), particularly for security applications. PSI [55] proposes a per-client, isolated virtual security appliance that provisions context-aware security function chains with software-defined network (SDN) capability. SecMANO [41] introduces a flexible NFV based security service system that can be orchestrated by the users. Recently, TrustAV [11] proposes malware detection outsourcing to untrusted environments by overcoming performance and memory restrictions of TEE.

**Verifiable network functions.** More recently, thanks to the wide availability of trusted hardware, outsourced network functions start to offer verifiability to remote clients. S-NFV [48], ClickSGX [9], and SafeBricks [42] provide integrity and confidentiality of the network function executions with hardware supported attestation, particularly by protecting codes and data in a protected memory area, called an enclave, of Intel SGX [32]. mbTLS [34] also exploits Intel SGX to enable middlebox operations on the plaintext packets in TLS-encrypted sessions. LightBox [13] further improves the flow processing inside a secure enclave to overcome its inherent performance bottleneck. Fritz et al.'s work [1] extends the verifiability of outsourced functions with Intel SGX to accounting capability.

## 8 CONCLUSION

While the security function outsourcing market is quickly growing to combat ever-evolving cyber threats, it has been difficult to utilize on-path transit networks, such as Tier-1 ISPs, for security outsourcing at full scale due to the lack of end-to-end protocols. Opsec is the first attempt to make on-path security outsourcing available to end users by automating the discovery and orchestration of user-requested security functions in transit networks. We hope our Opsec proposal helps in taking a step closer towards highly versatile on-path security function outsourcing in the current Internet.

---

[||]For our discussions, we consider that transit ISPs do not have much, if at all, direct end customers but mainly sell transit service to other ISPs; e.g., Tier-1 ISP.

## ACKNOWLEDGEMENT

This research is supported by the National Research Foundation, Prime Minister's Office, Singapore under its Corporate Laboratory@University Scheme, National University of Singapore, and Singapore Telecommunications Ltd.

# APPENDIX

## A  Integer linear programming for the optimal routing of Opsec traffic

Here we solve the problem of finding the optimal paths in an ISP network that has the demands of both legacy traffic and Opsec traffic. In comparison to legacy traffic, Opsec traffic is additionally constrained to visit a router connected to an Opsec-box, henceforth referred to as an Opsec-box node. Finding the shortest paths for legacy traffic is a traditional problem; and we observe that solving for Opsec traffic is similar to solving the traditional problem by splitting a traffic flow into two — one from source to Opsec-box nodes and another from Opsec-box nodes to target node. We first present the problem formulation for the case where an ISP has only one Opsec-box node, and subsequently extend it to the case of multiple Opsec-box nodes.

**Problem description.** Let $\mathcal{G} = \{V, \mathcal{E}\}$ denote an ISP's network, with $V$ being the set of routers (or nodes) and $\mathcal{E}$ the set of links connecting nodes in $V$. That is, $\mathcal{E} = \{(v, w) | v, w \in V\}$. We use $V^x, V^x \subset V$, to represent the set of source and target nodes of the network. Additionally, we define the following:

- $V^i$ : $V \setminus V^x$, the set of internal nodes;
- $m$ : $m \in V^i$ is an Opsec-box node;
- $c_{v,w}$ : cost of a link $(u, w) \in \mathcal{E}$;
- $\mathcal{T}^L_{s,t}$ : legacy traffic demand from node $s \in V^x$ to $t \in V^x$;
- $\mathcal{T}^P_{s,t}$ : Opsec traffic demand from node $s \in V^x$ to $t \in V^x$;
- $\mathcal{L}^{s,t}_{v,w}$ : legacy traffic volume of the flow from $s \in V^x$; to $t \in V^x$ routed on the link $(v, w) \in \mathcal{E}$;
- $\mathcal{P}^{s,t}_{v,w}$ : Opsec traffic volume of the flow from $s \in V$ to $t \in V$ routed on the link $(v, w) \in \mathcal{E}$.

Given the input traffic demands, $\mathcal{T}^L$ and $\mathcal{T}^P$, the problem is to find a cost-optimal set of paths to route traffic flows in the network $\mathcal{G}$. The total cost of routing is defined as the sum of the cost of all links in $\mathcal{E}$ carrying traffic, under the common assumption that the cost of a link is proportional to the traffic carried by it. We define the objective function for the optimization problem:

$$\text{Minimize} \sum_{(s,t) | s,t \in V^x} \sum_{(v,w) \in \mathcal{E}} (\mathcal{L}^{s,t}_{v,w} + \mathcal{P}^{s,t}_{v,w}) \times c_{v,w} \quad (1)$$

where $\mathcal{L}^{s,t}_{v,w}$ and $\mathcal{P}^{s,t}_{v,w}$ are the solution variables. That is, the matrix $\mathcal{L}^{s,t}_{v,w}$ ($\mathcal{P}^{s,t}_{v,w}$) denotes the legacy traffic (Opsec traffic) between source node $s$ and target node $t$ routed on link $(u, v) \in \mathcal{E}$. Next, in order to define the constraints, we list down the well-known constraints for finding the shortest path to route legacy traffic $\mathcal{T}^L$ over $\mathcal{G}$:

$$\text{Constraints}(\mathcal{L}, \mathcal{T}^L)$$
$$\sum_{y \in V} \mathcal{L}^{s,t}_{v,y} - \sum_{y \in V} \mathcal{L}^{s,t}_{y,v} = \begin{cases} \mathcal{T}^L_{s,t} & \text{if } v \text{ is } s, \\ -\mathcal{T}^L_{s,t} & \text{if } v \text{ is } t, \\ 0 & \text{otherwise.} \end{cases} \quad (2)$$
$$\forall (s,t) \in V^x; \forall v \in V$$

We utilize the above set of constraints to solve the optimization problem. Observe that, the optimal path routing problem for Opsec traffic can be abstracted as two sub-problems: in the first sub-problem, we need to optimally route Opsec traffic from all source nodes to Opsec-box nodes, and in the second, we need to route the same Opsec traffic received at Opsec-box nodes to all corresponding target nodes. For this purpose, we represent an $(s, t)$ Opsec traffic demand $\mathcal{T}^P_{s,t}$ in two parts, $\mathcal{S}_{s,m}$ and $\mathcal{S}_{m,t}$, and then apply the same set of constraints for legacy traffic on these two traffic demands. Algorithm 1 provides the steps for solving this problem in the case where there is only one Opsec-box node in the ISP network. Algorithm 1 takes four inputs, i.e., an ISP net-

---

**Algorithm 1** Optimal cost routing on Opsec adopted ISP

1: **Input:** $\mathcal{G} = \{V, \mathcal{E}\}; \mathcal{T}^L_{s,t}; \mathcal{T}^P_{s,t}; m$
2: $C \leftarrow \text{EmptyList}()$                         ▷ Initialize list of constraints
3: Add $\text{Constraint}(\mathcal{L}, \mathcal{T}^L)$ to $C$         ▷ for legacy traffic
4: **for** $(s, t) \in V^x$ **do**
5:     $\mathcal{S}_{s,m} \leftarrow \mathcal{T}^P_{s,t}$
6:     Add $\text{Constraint}(\mathcal{P}^{s,m}, \mathcal{S})$ to $C$     ▷ for Opsec traffic
7:     $\mathcal{S}_{m,t} \leftarrow \mathcal{T}^P_{s,t}$
8:     Add $\text{Constraint}(\mathcal{P}^{m,t}, \mathcal{S})$ to $C$     ▷ for Opsec traffic
9: **end for**
10: Set Equation 1 as objective
11: Solve ILP problem

---

work $\mathcal{G}$, legacy traffic matrix $\mathcal{T}^L$, Opsec traffic matrix $\mathcal{T}^P$, and the Opsec-box node $m$. First, it adds the constraints for the legacy traffic into the constraint set $C$. Then, as mentioned above, each Opsec traffic demand $\mathcal{T}^P_{s,t}$ is separated into two flows, such that one flow demand has the Opsec-box as its target and the other has the Opsec-box as its source node. $\mathcal{S}$ is a new traffic matrix having these separated traffic flows. The two corresponding set of constraints, $\text{Constraints}(\mathcal{P}^{s,m}, \mathcal{S})$ and $\text{Constraints}(\mathcal{P}^{m,t}, \mathcal{S})$ are added to $C$ for each $(s, t)$. Lastly, with the objective function given in Equation 1, we solve this integer linear programming (ILP) problem using a standard ILP solver (CPLEX [8]).

**Multiple Opsec-boxes.** If $M$ is the set of Opsec-box nodes in the ISP network $\mathcal{G}$, such that $m \in M$, then $M \subset V^i$. When a network has multiple Opsec-box nodes, Opsec traffic can be routed to multiple Opsec-box nodes; and therefore the traffic demand $(\mathcal{T}^P_{s,t})$ might not all go through a single Opsec-box node, but through multiple of them. This makes it different from the single Opsec-box scenario. To address this new requirement, we define a logical node $\overline{m}$ representing the set of Opsec-box nodes $M$; in addition, let $\overline{V}^x$ be the set of external nodes including $\overline{m}$; i.e., $\overline{V}^x = V^x \bigcup \{\overline{m}\}$.

With the above definitions, we replace Equation 2 with Equation 3 and Equation 4 below, for routing Opsec traffic.

$$\sum_{y \in \mathbf{V}} \mathcal{P}^{s,t}_{v,y} - \sum_{y \in \mathbf{V}} \mathcal{P}^{s,t}_{y,v} = \begin{cases} \mathcal{S}_{s,t} & \text{if } v \text{ is } s \\ -\mathcal{S}_{s,t} & \text{if } v \text{ is } t \\ 0 & \text{otherwise} \end{cases} \quad (3)$$

$$\forall (s,t) \in \mathbf{V}^{\mathbf{x}}, \forall v \in \mathbf{V} \setminus \mathbf{M};$$

$$\sum_{m \in \mathbf{M}} \sum_{y \in \mathbf{V}} (\mathcal{P}^{s,t}_{m,y} - \mathcal{P}^{s,t}_{y,m}) = \begin{cases} \mathcal{S}_{s,t} & \text{if } s \text{ is } \overline{m} \\ -\mathcal{S}_{s,t} & \text{if } t \text{ is } \overline{m} \\ 0 & \text{otherwise} \end{cases} \quad (4)$$

$$\forall (s,t) \in \mathbf{V}^{\mathbf{x}'}.$$

Equation 3 is the set of constraints for Opsec flow conservation on the nodes which do not have an Opsec-box, and Equation 4 ensures flow conservation on the Opsec-box nodes. Lastly, to address the flow conservation between two separated flows in $\mathcal{S}$, we add the following constraint:

$$\sum_{y \in \mathbf{V}} (\mathcal{P}^{s,\overline{m}}_{y,m} - \mathcal{P}^{s,\overline{m}}_{m,y} + \mathcal{P}^{\overline{m},t}_{y,m} - \mathcal{P}^{\overline{m},t}_{m,y}) = 0 \quad (5)$$

$$\forall m \in \mathbf{M}; \forall (s,t) \in \mathbf{V}^{\mathbf{x}}.$$

## B Security Analysis

In this section, We analyze potential threats from adversaries to Opsec components, i.e., client, server, and ISPs on the path, who may exploit Opsec protocol or harm the system itself for malicious purposes. Against the threats, we show how our design is robust with corresponding countermeasures.

*B.1 Dishonest/Malicious ISPs.* Opsec builds the chain of trust for the arbitrary and untrusted service providers with trusted execution environments (TEEs). The honest but curious, and even malicious outsourcing providers have been an important security concern of the middlebox applications, not limited to the security function outsourcing. The introduction of TEE contributes to solving critical security issues in many middlebox architectures, such as, [1, 20, 42]. Opsec, as well, addresses the following security problems by using TEE-adopted middlebox.

**Cheating by Dishonest ISPs.** A dishonest on-path ISP may attempt to execute a modified security function SF′ instead of the promised SF for economic gain (e.g., using less compute resources for the promised tasks). Since Opsec relies on TEEs, a dishonest ISP, however, cannot compromise the integrity property of TEE enclaves, which is protected with the remote attestation infrastructure, such as Intel Attestation Service [21]. The remote attestation system verifies the processes of asymmetric key generation, dec/encryption and signing, instead of verifying the public key through the PKI. TEE ensures the confidentiality of the symmetric and asymmetric keys and secures the decrypted payload only in the enclave.

**Denial of Service Attacks by Disruptive ISP.** On the other hand, any ISPs can exploit their networking capabilities to disrupt any protocols passing their network as well as Opsec. When an ISP attempts to disrupt Opsec service targeting a client or the next ISP who may want to serve a security function to the client without regarding loss of the connection, there are few countermeasures except to refrain to routing to the disruptive ISP. As similar to the typical DoS and ransom attacks, a disruptive ISP may earn economic benefits by attacking the availability of service or by leading the potential victims to avoid to use Opsec services, however, the disruptive ISP is a very strong assumption because it is highly risky and expensive to an ISP compared to the potential benefits one can get by the disruption.

*B.2 Adversaries at End-points.* Adversaries at the end-points of a connection, i.e., client and server, may attempt to exploit Opsec system to make benefits.

**Disruption by a malicious server.** A strong adversary may have direct control over an HTTP(S) server and attempt to identify and disrupt and/or exploit the requests of the Opsec services based on their featured HTTP messages. However, the Opsec protocol is robust against such attacks except shutting down its connection entirely that leads to the loss of its potential victim by preventing a user from accessing the malicious contents. While Opsec messages from a client are read by Opsec-ISPs before a server disrupts the request, the crafted Opsec response message do not influence the ISPs on the downstream path since the ISPs do not need to read or refer to the response messages from the previous-hops.

**Fake-Opsec ISP.** If the malicious server with TEE acts as a (fake) Opsec ISP, the server also has to follow the Opsec service request process and has to provide the offered service correctly as the dishonest ISPs have to do so. Another possible threat from a fake-Opsec ISP is overcharging. By intentionally responding a huge garbage content from a web server under the control, the fake-Opsec ISP may attempt to inflate its service expense. However, the volumetric payload from a fixed peer, i.e., the web server, is easily detectable by existing solutions.

**Protocol exploiting denial of service attacks.** An adversary may attempt to overload the Opsec-ISP with a denial-of-service; e.g., flooding Opsec requests, or creating half-opened requests. In this case, a well-known solution can be used. When such a DoS attack is detected, Opsec-box can request all the users to solve *puzzles* [40] to get prioritized service. This way, we can effectively restrict the malicious incoming traffic rate. Another possible exploitation is to use the attached Opsec responses by Opsec-ISPs during the handshake for amplification attack targeting a web-server. Because the amplification by the response messages only occurs during the handshake, Opsec request flooding to an ISP should precede the amplification attack to a web-server, that can be detected and handled by an Opsec-ISP.